\documentclass[prl,superscriptaddress,reprint,amsfonts,amssymb,showpacs]{revtex4-1}
\pdfoutput=1
\usepackage{graphics}
\usepackage{bm}
\usepackage{hyperref}

\begin{document}

\title{
Hybridizing ferromagnetic magnons and microwave photons in the quantum limit
}
\author{Yutaka Tabuchi}
\email{tabuchi@qc.rcast.u-tokyo.ac.jp}
\author{Seiichiro Ishino}
\author{Toyofumi Ishikawa}
\author{Rekishu Yamazaki}
\author{Koji Usami}
\affiliation{Research Center for Advanced Science and Technology (RCAST), The University of Tokyo, Meguro-ku, Tokyo 153-8904, Japan}
\author{Yasunobu Nakamura}
\affiliation{Research Center for Advanced Science and Technology (RCAST), The University of Tokyo, Meguro-ku, Tokyo 153-8904, Japan}
\affiliation{Center for Emergent Matter Science (CEMS), RIKEN, Wako, Saitama 351-0198, Japan}

\date{\today}

\begin{abstract}
We demonstrate large normal-mode splitting between a magnetostatic mode (the Kittel mode) in a ferromagnetic sphere of yttrium iron garnet and a microwave cavity mode. Strong coupling is achieved in the quantum regime where the average number of thermally or externally excited magnons and photons is less than one. We also confirm that the coupling strength is proportional to the square root of the number of spins. A nonmonotonic temperature dependence of the Kittel-mode linewidth is observed below 1 K and is attributed to the dissipation due to the coupling with a bath of two-level systems. 
\end{abstract}

\pacs{
03.67.Lx, 
42.50.Pq, 
75.30.Ds, 
76.50.+g  
}

{\maketitle}

Coherent coupling between spin ensembles and superconducting quantum circuits is now widely studied for implementations of quantum memories and transducers~\cite{bib:Imamoglu09,bib:Kubo10,bib:Amsuess11,bib:Zhu11,bib:Kubo11,bib:Saito13,bib:Schuster10,bib:Bushev11,bib:Probst13,bib:Ranjan13,bib:Wesenberg09,bib:Morton10,bib:Cecile14}. While the interaction of a single spin to the electromagnetic field mode is extremely weak, collective enhancement, which scales with the square root of the total number of spins, makes the effective coupling strong enough as an ensemble~\cite{bib:Imamoglu09}. Recent works on ensemble of nitrogen-vacancy centers in diamond have demonstrated coherent interaction between spin excitations and superconducting resonators~\cite{bib:Kubo10,bib:Amsuess11} and qubits~\cite{bib:Kubo11,bib:Zhu11,bib:Saito13}. Spin ensembles in other materials, such as rare-earth doped oxides, have also been investigated~\cite{bib:Schuster10,bib:Bushev11,bib:Probst13,bib:Ranjan13}. Multimode quantum memories utilizing a number of degenerate spatial modes are targeted in those paramagnetic spins~\cite{bib:Wesenberg09,bib:Morton10,bib:Cecile14}, while the same property causes vulnerability to spatial-mode mismatch with the coupling field. Moreover, the trade-off in the spin density remains a challenge for these systems: higher spin density allows stronger coupling with the electromagnetic field, while spin-spin interactions among the ensemble shorten the coherence time drastically with increasing spin density. 

In this context, ferromagnets represent a radically different limit. They typically have much higher density of spins which are strongly interacting and correlated. Due to the strong exchange interaction in between, spins are perfectly ordered in the ground state, and the low-lying excitations are collective waves of small-angle spin precession. In a sample with finite dimensions, dipolar interaction dominates the long-wavelength limit, and the boundary condition at the surface defines rigid discrete modes called magnetostatic modes. For spheroids, Walker analytically derived the energies and the spatial distributions of the modes~\cite{bib:Walker58,bib:Fletcher59}. In particular, the spatially uniform mode is called Kittel mode, which entails a huge magnetic dipole moment. Coherent excitation of the magnetostatic modes with electromagnetic field has been intensively studied in ferromagnetic resonance (FMR). However, the quantum limit to the level of single-magnon excitation has not been investigated. The goal in this Letter is to reveal the properties in the unexplored limit and demonstrate the potential of the hybrid system for applications in quantum information science. Quantum engineering of a large number of spins in a rigidly extended collective mode is analogous with optoelectromechanics using mechanical eigenmodes in nanostructures, now widely studied as one of the promising hybrid quantum technologies~\cite{bib:Aspelmeyer13}.

The ferromagnetic material we focus on is yttrium iron garnet (YIG). YIG is a representative {\it ferri}\,-magnetic insulator with the Curie temperature of about 550~K, and is commonly used for microwave components such as oscillators and band-pass filters. YIG in the ordered phase has a net spin density of $2.1 \times 10^{22}~\mu_{\rm B} $~cm$^{-3}$ ($\mu_{\rm B}$; Bohr magneton), orders of magnitude higher than  $10^{16}$-$10^{18}$~$\mu_{\rm B}$~cm$^{-3}$ in paramagnetic spin ensembles used in quantum memory experiments. Being a highly insulating material, it is free from dissipation due to conduction electrons, which results in a narrow-linewidth in FMR. The long lifetime of collective spin excitations in YIG has also attracted a lot of interests in the emerging field of spintronics~\cite{bib:Uchida10,bib:Kajiwara10}. It is of great interest to extend the scope of the magnon engineering to the quantum regime.

In this work, we demonstrate strong coupling of the Kittel mode in an undoped single-crystal YIG sphere to a microwave mode in a three-dimensional rectangular cavity. We have chosen spherical samples to avoid effects of inhomogeneous demagnetization field. %
Our result is to be compared with the prior work using a rectangular crystal placed on a superconducting coplanar waveguide (CPW) resonator \cite{bib:Huebl13}: (i) There existed inhomogeneities in the static field due to the demagnetization field and in the microwave field around the narrow CPW resonator, which could couple many other magnetostatic modes to the resonator mode. (ii) The broad magnon linewidth due to the Ga-doped YIG crystal prevented them from resolving the spectral lines at the anticrossing. (iii) The limited filtering in the microwave measurement setup allowed thermal excitations of magnons and photons in the sample. The present work resolves all these issues as discussed below. Nearly uniform microwave field in the large cavity volume, combined with the selection of a spherical sample, suppresses the coupling to other magnetostatic modes because of their symmetry. Meanwhile, the high spin density of YIG and the large volume of the sphere lead to enormous magnetic-dipole coupling strength between the Kittel mode and the cavity mode, namely, two harmonic oscillators. Normal-mode splitting with the cavity mode is clearly observed even in the quantum regime where both the average numbers of thermally excited magnons and photons are nearly zero and that of the probe microwave photons in the cavity is less than one.

\begin{figure}
  \centering
  \includegraphics{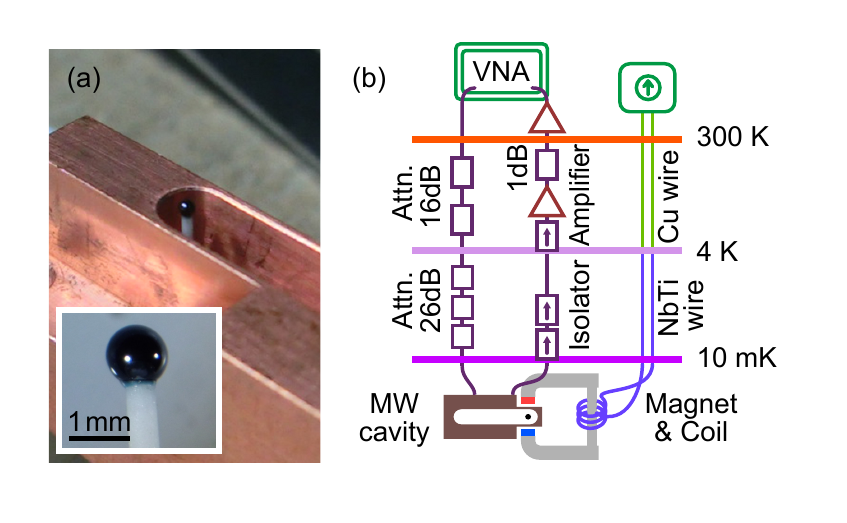}
  \caption{\label{fig:setup} (Color online) Experimental setup. (a) YIG sphere mounted in a rectangular cavity made of oxygen free copper. The sphere is glued to an alumina (aluminum-oxide) rod oriented to the crystal axis $\langle$110$\rangle$. The inset is a magnified picture of a sphere with a diameter of 1.0~mm. The cavity has two connector ports for transmission spectroscopy and dimensions of 22$\times$18$\times$3 mm that give the fundamental-mode (TE${}_{101}$) resonant frequency $\omega_{\rm c}/2\pi$ of 10.565~GHz. (b)~Measurement apparatus. The YIG sphere, cavity and magnet are cooled to 10~mK using a dilution refrigerator. A series of attenuators and isolators prevent thermal noise from reaching the sample space. The total attenuation of the input port is 48~dB at 10~GHz. The output signal is amplified by two low-noise amplifiers at 4~K and the room temperature. We use a vector network analyzer (VNA) for transmission and reflection spectroscopy. A static magnetic field perpendicular to the microwave magnetic field and parallel to the crystal axis $\langle$100$\rangle$ is applied by using permanent neodymium magnets and a magnetic yoke made of pure iron. We use a superconducting coil for fine tuning of the static field.}
\end{figure}

Our experimental setup is shown in Fig.~\ref{fig:setup}. The cavity made of oxygen free copper has the fundamental-mode (TE${}_{101}$) frequency $\omega_{\rm c}/2\pi$ of 10.565~GHz, and its internal cavity loss $\kappa_{\rm int}/2\pi$ is about~1.0~MHz at low temperature. This cavity has two connector ports for the transmission spectroscopy; an asymmetric port configuration is used where input and output ports have different coupling strengths $\kappa_1/2\pi$ and $\kappa_2/2\pi$ of 0.13 and 1.5~MHz, respectively. A YIG sphere made by Ferrisphere Inc.~\footnote{\url{http://www.ferrisphere.com/}} is mounted in the cavity at the magnetic antinode of the fundamental mode. We apply a static magnetic field of approximately 370~mT along the crystal axis $\langle$100$\rangle$ which is the hard magnetization axis for YIG. The sample is supposed to be uniformly magnetized and saturated.

\begin{figure*}
  \centering
  \includegraphics{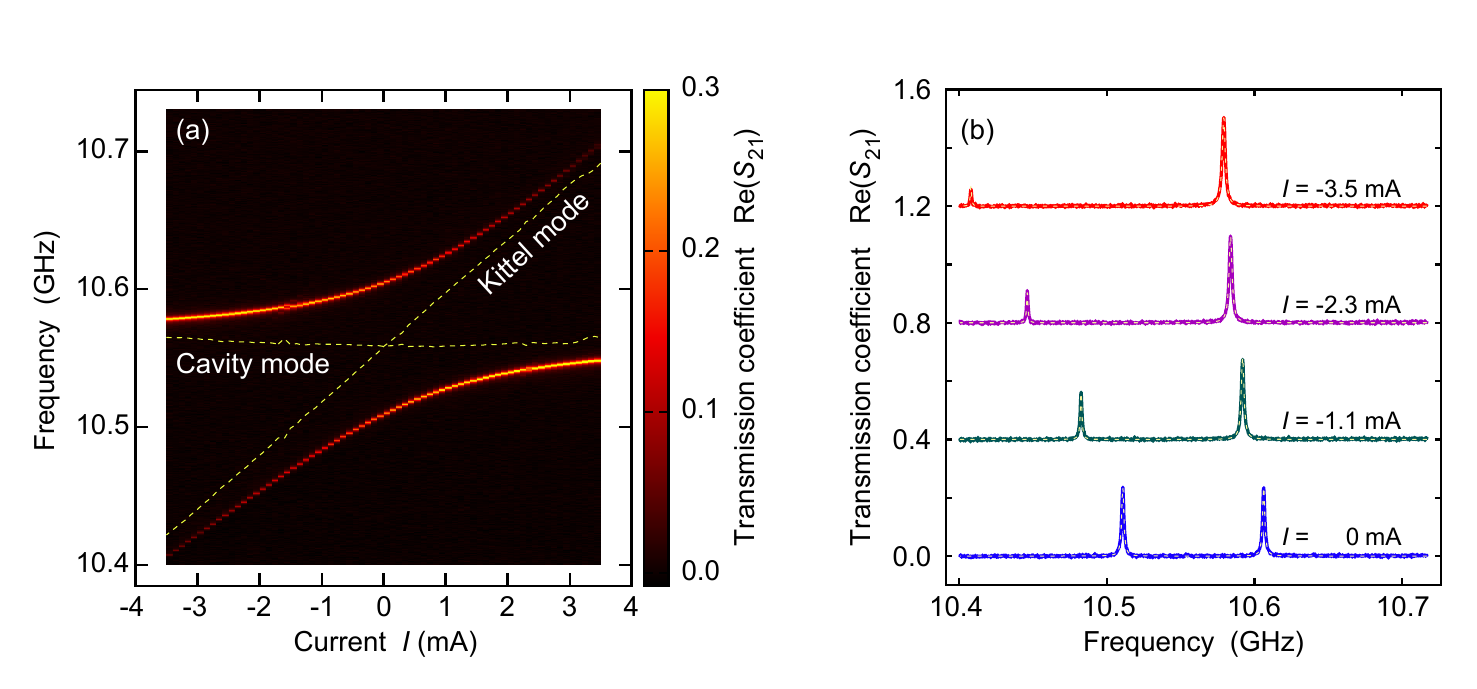}
  \caption{\label{fig:splitting} (Color online) Normal-mode splitting between the Kittel mode and the cavity mode TE$_{101}$. (a) Amplitude of the transmission Re($S_{21}$) through the cavity as a function of the probe microwave frequency and the static magnetic field presented in the current $I$ through the superconducting coil. The field-to-current conversion ratio obtained by fitting the FMR frequency is 1.42~mT/mA. The phase offset in $S_{21}$ is adjusted so that the peak has a pure absorption spectrum. The horizontal dashed line shows the cavity resonant frequency, while the diagonal dashed line shows the Kittel-mode frequency, both obtained from the fitting based on the input-output theory. We used a probe microwave power of $-123$~dBm, which corresponds to the average photon number of $0.9$ in the cavity. (b) Cross sections at static magnetic fields corresponding to $I$ = $-3.5$, $-2.3$, $-1.1$, and 0 mA. Solid curves are experimental data with vertical offset for clarity, and dashed white lines are the fitting curves.}
\end{figure*}

We first measure the transmission spectrum of the cavity loaded by a YIG sphere with a diameter of 0.5 mm. Figure~\ref{fig:splitting}(a) shows the transmission coefficient Re$(S_{21})$ as a function of the frequency and the magnetic field tuned by the bias current $I$ in the coil. 
A pronounced normal-mode splitting is observed, indicating strong coupling between a collective excitation mode in the YIG sphere and the cavity mode. We assign the mode in the sphere to the Kittel mode; it gives the maximum coupling strength to the nearly uniform cavity field and the frequency which is linearly dependent on the static field. Although it is difficult to recognize in Fig.~\ref{fig:splitting}(a), we also see a few hints of other tiny anticrossings, for example, at $I = -1.6$, $2.3$, and 3.3~mA. These are due to weak coupling of the cavity mode with other magnetostatic modes in the YIG sphere, induced by the small inhomogeneity of the magnetic fields. Several cross-sections of the 2D color plot are depicted in Fig.~2(b). As a function of the magnetic field, the Kittel mode approaches the cavity mode. At the degeneracy point ($I$ $\equiv$ 0~mA) where the Kittel-mode frequency coincides with the cavity frequency, we see the normal-mode splitting of nearly 100~MHz, orders of magnitude wider than the linewidths. At this point, the Kittel and cavity modes form `magnon-polariton' modes, i.e., hybridized modes between the collective spin excitation and the cavity excitation. 

We now evaluate the coupling strength and the cavity and magnon linewidths by fitting the transmission coefficient $S_{21}(\omega)$ with an equation derived from the input-output theory. The transmission coefficient of the hybrid system is written as
\begin{eqnarray}
 & & S_{21}(\omega) \nonumber \\
 & & = \frac{\sqrt{\kappa_1\kappa_2}}  {i(\omega-\omega_{\rm c})-\frac{\kappa_1+\kappa_2+\kappa_{\rm int}}{2}
  +\frac{|g_m|^2}{i(\omega-\omega_{\rm FMR})-\gamma_{\rm m}/2}},
  \label{eq:input-output}
\end{eqnarray}
where $g_{\rm m}$ is the coupling strength of the Kittel mode to the cavity, and $\omega_{\rm FMR}$ and $\gamma_{\rm m}$ are the frequency and the linewidth of the Kittel mode, respectively. The input and output port couplings of the cavity are determined separately by measuring additionally the reflection of the cavity output. As depicted by the dashed white lines in Fig.~\ref{fig:splitting}(b), the spectra are well fitted with Eq.~(\ref{eq:input-output}). The coupling strength of the Kittel mode to the cavity $g_{\rm m}/2\pi$, the total cavity linewidth $\kappa/2\pi=(\kappa_1+\kappa_2+\kappa_{\rm int})/2\pi$, and the Kittel-mode linewidth $\gamma_{\rm m}/2\pi$ are determined as 47~MHz, 2.7~MHz, and 1.1~MHz, respectively. From the parameters, the hybrid system turns out to be deep in the strong coupling regime where $g_{\rm m} \gg \gamma_{\rm m}, \kappa$, even at the lowest temperature and with the weakest probe power. A dimensionless measure which indicates how well the spins couple to the cavity mode is the cooperativity defined as $C = 4g^2_{\rm m}/\kappa\gamma_{\rm m}$. The obtained cooperativity is $C$ = $3.0 \times 10^3$, which is extremely large compared to the numbers achieved with paramagnetic spin ensembles.

\begin{figure}
  \centering
  \includegraphics{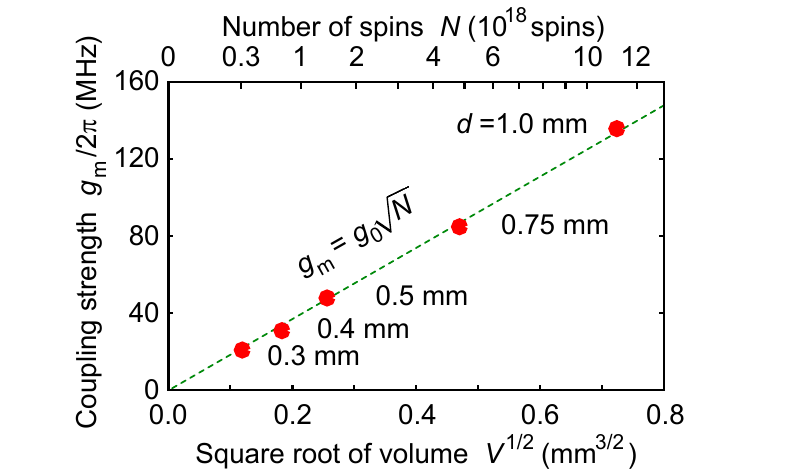}
  \caption{\label{fig:coupling} (Color online) Coupling strength of the Kittel mode to the microwave cavity mode as a function of the sample diameter $d$. The bottom and top horizontal axes show the square root of the sphere volume and the corresponding net spin numbers, respectively. The dashed line is a linear fit. The slope of the line gives the single-spin coupling strength $g_0/2\pi$ to the cavity mode, which is estimated to be 39~mHz.}
\end{figure}

We also check dependence of the coupling strength on the sphere diameter. 
Provided that all the net spins in the ferrimagnetic YIG sphere are precessing in phase, the coupling strength $g_{\rm m}$ of the Kittel mode to the cavity mode is expected to be proportional to the square root of the number of the net spins $N$, i.e., $g_{\rm m} = g_{\rm 0} \sqrt{N}$ where $g_{\rm 0}$ is the coupling strength of a single Bohr magneton to the cavity. The single-spin coupling strength is calculated to be $g_{\rm 0}/2\pi = \gamma_{\rm e} \sqrt{\mu_0 \hbar\omega_{\rm c}/V_{\rm c}}/2\pi$ = 38~mHz for TE${}_{101}$ mode, where $\gamma_{\rm e}$ is the electron gyromagnetic ratio of $2\pi$$\times$28.0~GHz/T, $\mu_0$ is the permeability of vacuum, and $V_{\rm c}$ is the volume of the cavity. The enhancement by the factor of $\sqrt{N}$ is due to a magnon excitation in the Kittel mode, i.e., constructive interference of all possible processes in which a cavity photon flips one of the spins in the sphere. The factor $\sqrt{N}$ naturally appears as the Clebsch-Gordan coefficient when the raising operator of the total spin is applied to a fully polarized $N$-spin system~\cite{bib:Agarwal84}. The red dots in Fig.~\ref{fig:coupling} show the obtained coupling strength as a function of the sphere diameter. It clearly demonstrates that the coupling strength is proportional to the square root of the number of spins. We evaluate the single-spin coupling strength $g_0/2\pi$ to be 39~mHz from the fitting. The good agreement with theory indicates that the design of our hybrid system is reliable and robust.

\begin{figure}
  \includegraphics{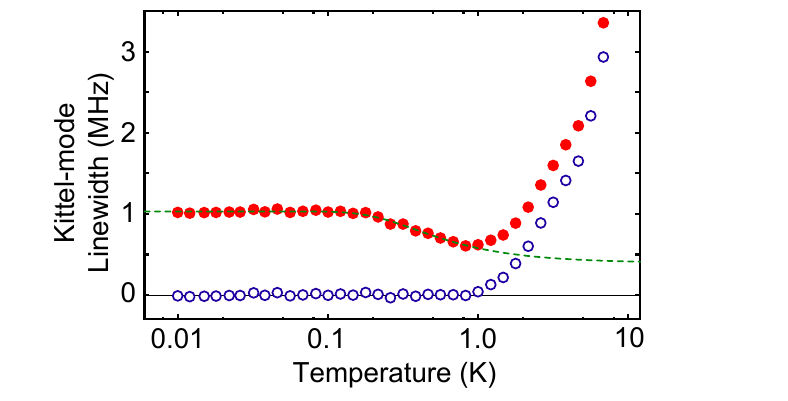}
  \caption{(Color online) Linewidth of the Kittel mode in the 0.5-mm sample as a function of the temperature. The red dots show the linewidth obtained by fitting $S_{21}(\omega)$ measured at each temperature. The dashed line is the fitting curve to the temperature dependence below 1~K. This is based on a model in which the linewidth is dominated both by transverse relaxation to two-level systems (TLS) near-resonant with the Kittel-mode and by surface scattering of the Kittel-mode magnons into other degenerate modes. The blue circles show the deviation from the model. The line broadening above 1~K is ascribed to the slow-relaxation process due to impurity ions and magnon-phonon scattering. \label{fig:linewidth_td}}
\end{figure}

We further observe a peculiar temperature dependence of the Kittel-mode linewidth below 1~K; little has been known about the dependence in this temperature range~\cite{bib:Gurevich96}. The red dots in Fig.~\ref{fig:linewidth_td} show the Kittel-mode linewidth as a function of the temperature. As seen also in Refs.~\citenum{bib:Spencer59} and \citenum{bib:Spencer61}, in the temperature range from 10~K to 1~K, the linewidth monotonically decreases. The dominant mechanisms of the relaxation here are known to be the so-called slow-relaxation due to impurity ions~\cite{bib:Teale62} and magnon-phonon scattering~\cite{bib:Sparks61}. The observed linewidth below 1~K, however, increases as temperature decreases. Such behavior has been predicted to be a signature of the transverse relaxation due to two-level systems (TLSs)~\cite{bib:Vleck64}, but has never been observed in FMR. Note that an analogous behavior has been seen, for example, in superconducting coplanar waveguide resonators~\cite{bib:Jiansong08}. The mechanism is explained as follows: The magnon excitation in the Kittel mode decays into an ensemble of near-resonant TLSs. As the relaxation rate to the TLSs is proportional to their polarization, the linewidth of the Kittel mode is expected to be $\gamma_{\text{TLS}}(T) = \gamma_{\text{TLS}}(0) \tanh(\hbar\omega_{\rm FMR}/2k_{\rm B} T)$, where $\gamma_{\text{TLS}}(0)$ is a constant giving the low-temperature value depending on the spectral properties of the bath of TLSs and $k_{\rm B}$ is the Boltzmann factor. In the analysis of the data, we also take into account a temperature-independent contribution, $\gamma_{\text{mm}}$, of the elastic intermode magnon-to-magnon scattering at surface~\cite{bib:Sparks61}. The fitting curve $\gamma_{\text{m}}(T) = \gamma_{\text{TLS}}(T) + \gamma_{\text{mm}}$ shown as the dashed line in Fig.~\ref{fig:linewidth_td} well fits to the experimental data below 1~K. The fractions of the Kittel-mode linewidth ascribed to the TLSs and the surface scattering $\gamma_{\text{TLS}}(0)/2\pi$ and $\gamma_{\text{mm}}/2\pi$ are estimated to be 0.63~MHz and 0.39~MHz, respectively. The latter linewidth arisen from the surface scattering is consistent with that in Ref.~\citenum{bib:Spencer59} at 4.2~K. The line-broadening due to TLSs can be observed only when the system is in the quantum regime, where the Kittel mode as well as the TLSs is cooled down toward the ground states. Note that these contributions to the linewidth are not likely intrinsic ones. A loss measurement with the parallel pumping method implied much narrower linewidth of about 45~kHz at 4.2~K~\cite{bib:Spencer61}. Improvement of the sample quality, e.g., in surface roughness and densities of impurities and defects, would be beneficial for future development toward coherent control of magnon excitations.

To conclude, we achieved strong coupling of a ferromagnetic magnon mode and a microwave cavity mode in the quantum regime where the number of average excitations in the hybrid system is less than one. The obtained coupling strength and the linewidths of the Kittel and the cavity modes are 47~MHz, 1.1~MHz, and 2.7~MHz, respectively, deeply in the strong coupling regime with the cooperativity of 3.0$\times$$10^3$. We also confirmed the linear scaling of the coupling strength with the sample volume in accordance with the theory. %
Furthermore, the linewidth of the Kittel mode showed a peculiar temperature dependence at the lowest temperature, which is attributed to the decay of the magnon excitations to a bath of TLSs. 

A large spin density in ferromagnets easily couples the collective excitation to a cavity field even for a relatively large mode volume. This opens a path toward new hybrid systems between the magnetostatic modes in ferromagnets and superconducting quantum circuits. By mediating the interaction between the magnon modes and well-controlled superconducting qubits via a microwave cavity mode, generation and characterization of non-classical magnon states, such as Fock states and squeezed states, would be possible. 

{\it Note added}.---
Recently, a paper by Zhang \textit{et al}. appeared~\cite{bib:Zhang14}. With a similar setup, but at room temperature, they demonstrated phenomena such as magnetically induced transparency, the Purcell effect, etc. It would be of great interest to bring these intriguing physics to the quantum regime as well.

\begin{acknowledgments}
This work was partly supported by the Funding Program for World-Leading Innovative R\&D on Science and Technology (FIRST), the Project for Developing Innovation System of MEXT, JSPS KAKENHI (Grant Number 26600071), the Murata Science Foundation, Research Foundation for Opto-Science and Technology, and National Institute of Information and Communications Technology (NICT). 
\end{acknowledgments}


%

\end{document}